\documentclass[prb,aps,final,twocolumn,floatfix]{revtex4}
\input epsf
\begin{document}
\title{Theoretical study of finite temperature spectroscopy in van der
Waals clusters. II Time-dependent absorption spectra}
\author{F. Calvo, F. Spiegelman}
\affiliation{Laboratoire de Physique Quantique, IRSAMC, Universit\'e Paul
Sabatier, 118 Route de Narbonne, F31062 Toulouse Cedex, France}
\author{D. J. Wales}
\affiliation{University Chemical Laboratories, Cambridge CB2 1EW, United
Kingdom}
\begin{abstract}
Using approximate partition functions  and a master equation approach,
we  investigate  the  statistical  relaxation  toward  equilibrium  in
selected  CaAr$_n$   clusters.  The  Gaussian   theory  of  absorption
[J. Chem. Phys.,  {\bf 000}, 0000, 2003] is  employed to calculate the
average  photoabsorption   intensity  associated  with   the  $4s^2\to
4s^14p^1$  transition  of  calcium   as  a  function  of  time  during
relaxation. In CaAr$_6$ and  CaAr$_{10}$ simple relaxation is observed
with a  single time scale.  CaAr$_{13}$ exhibits much  slower dynamics
and the  relaxation occurs over two distinct  time scales. CaAr$_{37}$
shows much slower relaxation  with multiple transients, reminiscent of
glassy  behavior  due  to  competition  between  different  low-energy
structures.  We interpret  these results  in terms  of  the underlying
potential energy surfaces for these clusters.
\end{abstract}
\maketitle

\section{Introduction}

Clusters  of van der  Waals atoms  or molecules  are known  to exhibit
significant finite  size effects\cite{jortner} in  their thermodynamic
behavior.  In  particular, their structure  and low-temperature stable
phase can  strongly depend  on size.  By  doping such clusters  with a
chromophoric atom,  spectroscopic techniques are  able to characterize
the structure of  the whole complex.  Very often, the chromophore only
acts as a small perturbation on the geometry of the host van der Waals
cluster, and spectroscopy can then be used  as a probe  of structure and
structural changes  in the  cluster. These ideas,  which date  back to
experimental  studies in  the eighties,\cite{hahn,even}  have received
some theoretical support in the recent years.\cite{curotto,moseler} In
the  previous  paper in  this  series,\cite{paperI}  the influence  of
temperature  and  cluster  size  on the  photoabsorption  spectrum  in
CaAr$_n$ clusters was investigated.
In some  cases, such  as CaAr$_{13}$ or  CaAr$_{37}$, features  in the
absorption  spectrum  were  seen  to originate  from  the  competition
between  specific  isomers.  These  results  were  obtained using  the
quantum  superposition method,\cite{jcpquantum}  which  yields ergodic
data by construction from a  database of local minima on the potential
energy surface (PES).

However,  the equilibrium properties  computed from  the superposition
approach  obviously provide  no  information about  the  way in  which
equilibrium was established, or how long equilibration would take from
a  given   starting  distribution.   As   was  shown  by   Miller  and
coworkers,\cite{ar38miller}  relaxation can  be rather  slow  when the
competing  minima are separated  by large  (free) energy  barriers. In
cases such  as Ar$_{38}$, conventional  molecular dynamics simulations
are  presently unable  to reach  equilibrium and  thus  cannot provide
estimates  of  the  rate  constants.   However,  the  master  equation
dynamics  approach,   introduced  in  cluster  physics   by  Kunz  and
Berry,\cite{kunz}    has   been    successfully   applied    to   this
system,\cite{doyemw99a} although  a subsequent discrete  path sampling
study located somewhat faster pathways.\cite{dps} In particular, it has
been used to calculate activation energies in the interconversion of
(NaCl)$_{35}^-$ nanocrystals\cite{doyew99} that could be compared to
those extracted from ion mobility measurements.\cite{hudgins}

It is important to characterize  the dynamics of the CaAr$_n$ clusters
investigated  previously at  thermodynamical equilibrium,  because the
time required for reaching equilibrium may not always be accessible in
experiments. In  addition, the most powerful way  to investigate phase
changes  through  spectroscopy  on  single clusters  may  require  ion
trapping.\cite{maier,KruckebergSMP00} In principle, this technique can
be used to examine the system over very long time scales,  which makes
it a possible tool for studying dynamics, including  relaxation toward
thermal equilibrium, possibly via kinetic traps.

This article is a natural extension to our previous work, and provides
an investigation of the  relaxation dynamics in some selected CaAr$_n$
clusters, as  evidenced by their photoabsorption spectra.  In the next
section,  we  briefly  recall  the  basic ingredients  of  the  master
equation  approach,  and we  incorporate  quantum  corrections to  the
equilibrium  probabilities and rate  constants. These  corrections are
necessary  for  consistency  with the  previous  article.\cite{paperI}
Before actually studying dynamical processes, we construct and discuss
disconnectivity             graphs\cite{becker,walesacp}            in
Sec.~\ref{sec:landscapes}. Such graphs are very helpful in elucidating
the role of the underlying PES on the kinetics. Disconnectivity graphs
are presented  and discussed  for the clusters  CaAr$_6$, CaAr$_{10}$,
CaAr$_{13}$,  and CaAr$_{37}$,  which were  investigated  in reference
\onlinecite{paperI}. Using the  Gaussian theory of absorption inspired
by  Wadi and  Pollak,\cite{paperI,wp} we calculate the photoabsorption
intensity of these clusters as  a function of both time and excitation
energy.  The results  are given  and analysed  in Sec.~\ref{sec:tdas},
before we summarize and conclude in Sec.~\ref{sec:ccl}.

\section{Master equation dynamics}
\label{sec:med}

In  the master  equation approach,\cite{master,kunz95}  the interbasin
dynamics on the energy landscape is described by the time evolution of
a vector, ${\bf P}(t)$, whose  components are the probabilities of the
system residing  in each of the  basins at time  $t$. The differential
equations governing this evolution are
\begin{equation}
\frac{dP_i}{dt}    =    \sum_{j\neq    i}   \left[    k_{ij}    P_j(t)
-k_{ji}P_i(t)\right],
\label{eq:me1}
\end{equation}
where  $k_{ij}$ is  the  rate constant  for  transitions leading  from
minimum $j$ to  minimum $i$. A transition matrix  ${\bf W}$ is defined
with components\cite{master,kunz95,walesacp}
\begin{equation}
W_{ij} = k_{ij} - \delta_{ij}\sum_m k_{mi},
\label{eq:me2}
\end{equation}
and Eq.~(\ref{eq:me1}) is solved analytically after symmetrization and
diagonalization  of the  matrix  ${\bf W}$.\cite{kunz95,walesacp}  The
rate  constant  $k_{ij}$  is   the  sum  over  all  transition  states
separating minima $i$ and $j$:
\begin{equation}
k_{ij} = \sum_\alpha k_j^\alpha,
\label{eq:me3}
\end{equation}
where $k_j^\alpha$ is given by the usual Rice-Ramsperger-Kassel-Marcus
theory.    \cite{RiceR27,RiceR28,Kassel28,Marcus52,WiederM62,Marcus65}.
In the canonical ensemble we have
\begin{equation}
k_j^\alpha (T) = \frac{k_BT}{2\pi \hbar}\frac{Z_j^\alpha (T)}{Z_j(T)},
\label{eq:me4}
\end{equation}
where  $k_B$   and  $h$   are  Boltzmann's  and   Planck's  constants,
respectively.   $Z_j^\alpha$  and $Z_j$  are  the partition  functions
corresponding to  the transition state  $\alpha$ between $i$  and $j$,
and  to  the minimum  $j$,  respectively.  Here  we use  the  harmonic
approximation to model both $Z_j^\alpha$ and $Z_j$, but we incorporate
quantum   corrections  by  employing   the  expressions   for  quantum
oscillators:
\begin{eqnarray}
Z_j(T)&=&\frac{h_j}{(2\pi)^\nu N!}e^{-\beta E_j} \prod_{k=1}^\nu
\frac{\exp(-\beta\hbar               \omega_{jk}/2)}{1-\exp(-\beta\hbar
\omega_{jk})};
\label{eq:me5}\\
Z_j^\alpha(T)&=&\frac{h_j^\alpha}{(2\pi)^{\nu-1}N!}e^{-\beta E_j^\alpha}
\prod_{k=1}^{\nu-1}                              \frac{\exp(-\beta\hbar
\omega_{jk}^\alpha/2)}{1-\exp(-\beta\hbar \omega_{jk}^\alpha)}.
\label{eq:me6}
\end{eqnarray}
In these expressions, $h_j$  (resp. $h_j^\alpha$) denotes the order of
the  point  group of  minimum  $j$  (resp.  transition state  $\alpha$
between minima $i$ and $j$).  $E_j$ and $E_j^\alpha$ are the energy of
these    stationary    points,     and    $\{    \omega_{jk}\}$    and
$\{\omega_{jk}^\alpha\}$    are   their    respective    normal   mode
frequencies.  $\nu=3N-6$  is  the  number of  independent  degrees  of
freedom of the cluster, and $\beta=1/k_BT$. In the limit $\hbar\to 0$, the
rate constant tends to its usual classical value
\begin{equation}
k_j^\alpha(\hbar\to 0)     =    \frac{h_j^\alpha}{h_j}     \frac{(\bar
\omega_j)^\nu}{(\bar        \omega_j^\alpha)^{\nu-1}}        e^{-\beta
(E_j^\alpha-E_j)},
\label{eq:me7}
\end{equation}
where $\bar\omega_j$ and  $\bar\omega_j^\alpha$ are the geometric mean
normal  mode   frequencies  of  the  minimum   and  transition  state,
respectively.
In   the  low   temperature  limit   $\beta\to\infty$,  the   rate  is
proportional    to   $T\exp[-(E_j^{\alpha    0}-E_j^0)/k_BT]$,   where
$E_j^{\alpha 0}$  and $E_j^0$ are  the energies of the  two stationary
points   including   the   zero-point  energy   contributions.   While
$E_j^\alpha-E_j$ is always positive,  it may well be that $E_j^{\alpha
0}<E_j^0$, leading to  a divergent rate.  In this  case the two minima
cannot   be   treated   independently   because  of   strong   quantum
delocalization effects.   However, this situation should  not arise at
high or  moderate temperatures where  $k_BT$ is large compared  to the
zero-point energy, and  this is the regime of  interest in the present
work.

\begin{figure}[htb]
\vbox to 12cm{
\includegraphics{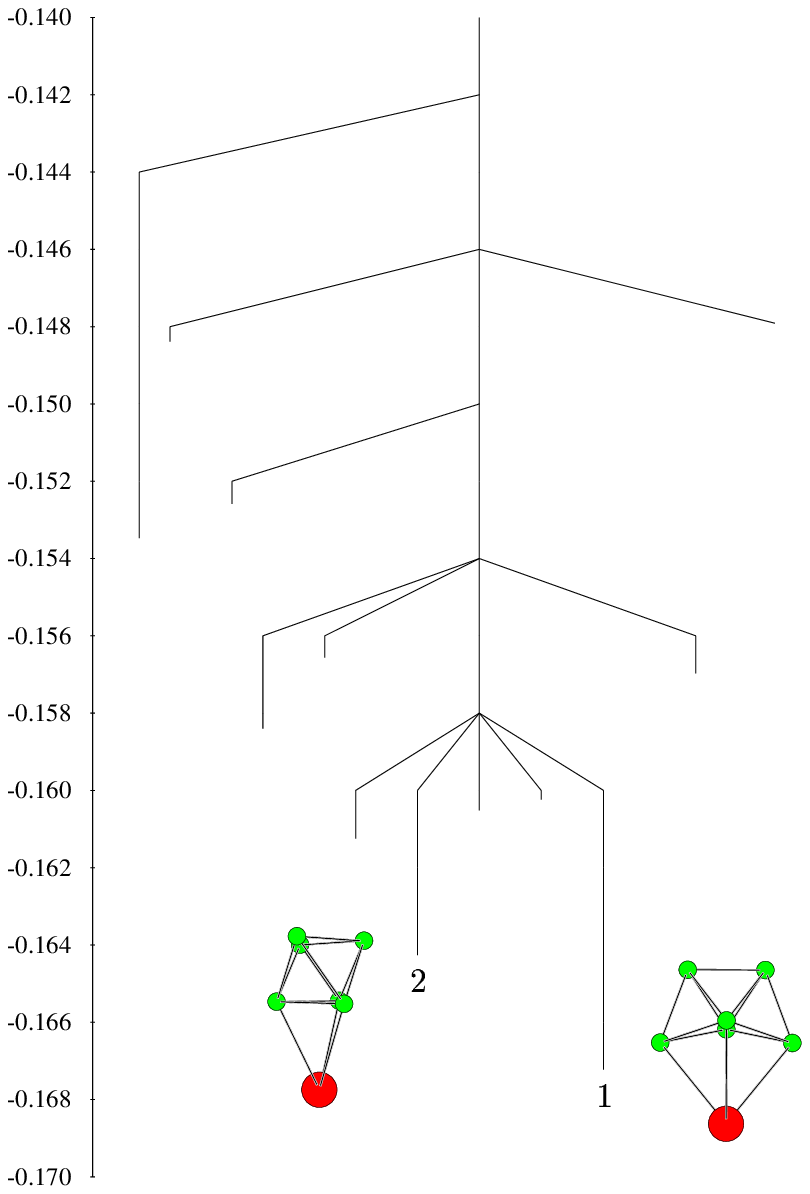}
\vfill}
\caption{Disconnectivity graph of the CaAr$_6$ cluster. The energies are in
eV.}
\label{fig:graph6}
\end{figure}
Application of  the master equation  approach requires one to  build a
connected    set   of    minima   and    the    interwell   transition
states.\cite{walesacp}       To       do       this      we       used
eigenvector-following\cite{ef} in conjunction  with several methods to
sample the local  minima.\cite{walesacp} For small clusters containing
less than 11  atoms, systematic searches were started  from the global
minimum, and continued until no  new minima were found after searching
for transition states  along every eigenvector of each  minimum in the
set. For  these systems the connected  set of minima  should be nearly
complete.  For CaAr$_{13}$,  we expect  the number  of  distinct local
minima  to be  significantly larger  than in  Ar$_{14}$,  which itself
possesses more  than $10^4$, excluding  permutation-inversion isomers.
We    decided   to    perform   only    one   cycle    of   systematic
eigenvector-following  transition  state searches  from  each mode  of
every minimum in the database collected for the study at thermodynamic
equilibrium.\cite{paperI}

The situation becomes quite critical  for CaAr$_{37}$. As was shown in
our  previous   study,\cite{paperI}  several  minima   with  different
structures coexist  at low  energy, and  we were not  able to  build a
connected  set  containing  all  these  minima  using  one  systematic
eigenvector-following cycle  from the database  obtained with parallel
tempering Monte  Carlo. For this  larger cluster, we selected  a small
set of  the low energy  minima of interest,  and we attempted  to find
discrete   pathways   (series   of  minima-transition   states-minima)
connecting  them.  To  do this  we  followed the  method described  in
reference \onlinecite{dps},  where a connected path is  built up using
successive  double-ended  pathway  searches.  The  nudged-elastic-band
approach
\cite{MillsJ94,JonnsonMJ98,HenkelmanUJ00,HenkelmanJ00,HenkelmanJ01,MaragakisABRK02}
was used to generate initial guesses for transition states, which were
then  used as  the  starting points  for hybrid  eigenvector-following
transition  state searches  \cite{munrow99,kumedamw01}.  These initial
paths generally  have rather high  barriers, and so the  discrete path
sampling  approach  was then  employed  to  provide  a more  realistic
account  of  the  rates.\cite{dps}  These  calculations  also  produce
connected  databases  of  minima  and transition  states  that  should
provide  a  more  appropriate  picture  of  the  dynamics,  either  in
visualisations using  disconnectivity graphs, or  from master equation
calculations.  All the geometry optimisations and pathway calculations
were performed  with the OPTIM  software package,\cite{optim,walesacp}
for which we implemented  the pairwise potential describing the ground
state potential energy surface of CaAr$_n$ clusters.\cite{epjd}

\section{Energy landscapes}
\label{sec:landscapes}

A  convenient  way  of  visualizing  complex energy  landscape  is  to
construct  a disconnectivity  graph.  Such graphs  were introduced  by
Becker and Karplus\cite{becker} in their study of a tetrapeptide. They
have  since   been  used  in  a   much  wider  context.\cite{walesacp}
Disconnectivity graphs show how the minima of the PES are connected to
each other and,  more importantly, what is the  height of the barriers
between  these  minima.  Further  details about  the  construction  of
disconnectivity     graphs    can     be     found    in     reference
\onlinecite{walesacp}.

In the present work we have included zero-point energy corrections for
the energies  of the  minima and transition  states.  Burnham  {\it et
al.\/}  have  previously  found  that zero-point  effects  change  the
appearance   of   disconnectivity  graphs   for   the  water   hexamer
significantly.\cite{BurnhamXMAM02} As  noted in the  previous section,
the  zero-point  terms sometimes  lead  to  the  merging of  two  (and
possibly  more) minima.   However, such  events were  not  observed in
CaAr$_6$,  for which  we  located twelve  distinct  minima and  twenty
transition  states.  The  disconnectivity  graph for  this cluster  is
shown  in  Fig.~\ref{fig:graph6}  along  with the  two  lowest  energy
isomers.  According   to  the  terminology   introduced  in  reference
\onlinecite{walesnat}, this graph has a typical `palm tree' character:
the global  minimum lies  at the  bottom of a  single funnel,  and the
absence of large energy barriers is expected to permit relatively fast
relaxation toward equilibrium.
\begin{figure}[htb]
\vbox to 12cm{
\includegraphics{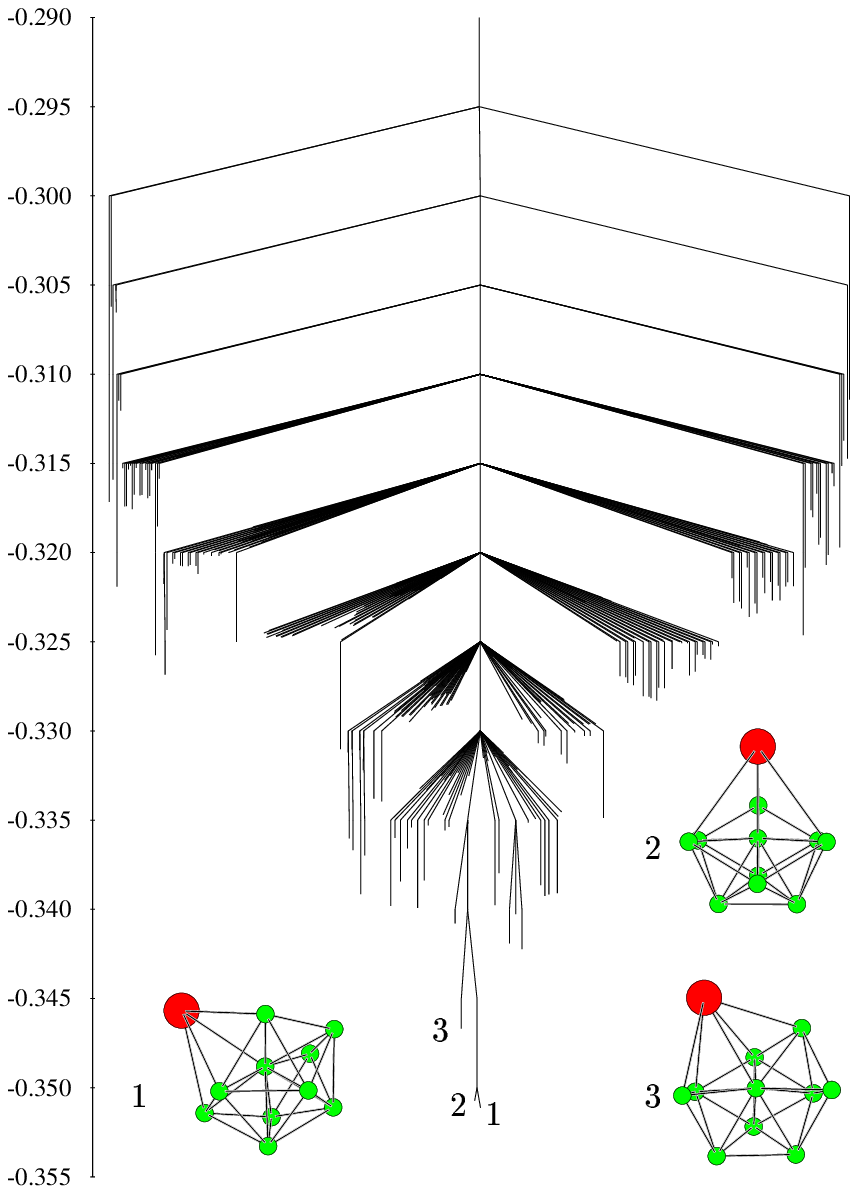}
\vfill}
\caption{Disconnectivity graph of the CaAr$_{10}$ cluster. The energies are in
eV.}
\label{fig:graph10}
\end{figure}

A similar  picture was found  for the larger cluster,  CaAr$_{10}$, as
shown in Fig.~\ref{fig:graph10}. In  this case we located 391 distinct
minima  connected via  1831 transition  states. At  the bottom  of the
funnel, the  three lowest-energy minima  lie close together.  They are
all based  on the  incomplete icosahedra, with  different substitional
sites for  the calcium atom. As Fig.~\ref{fig:graph10}  shows, the two
lowest minima  are extremely close  in energy, and the  energy barrier
that separates  them is  very small (less  than $10^{-3}$  eV).  Hence
these two isomers can barely be considered distinct.

\begin{figure*}[htb]
\vbox to 12cm{
\includegraphics{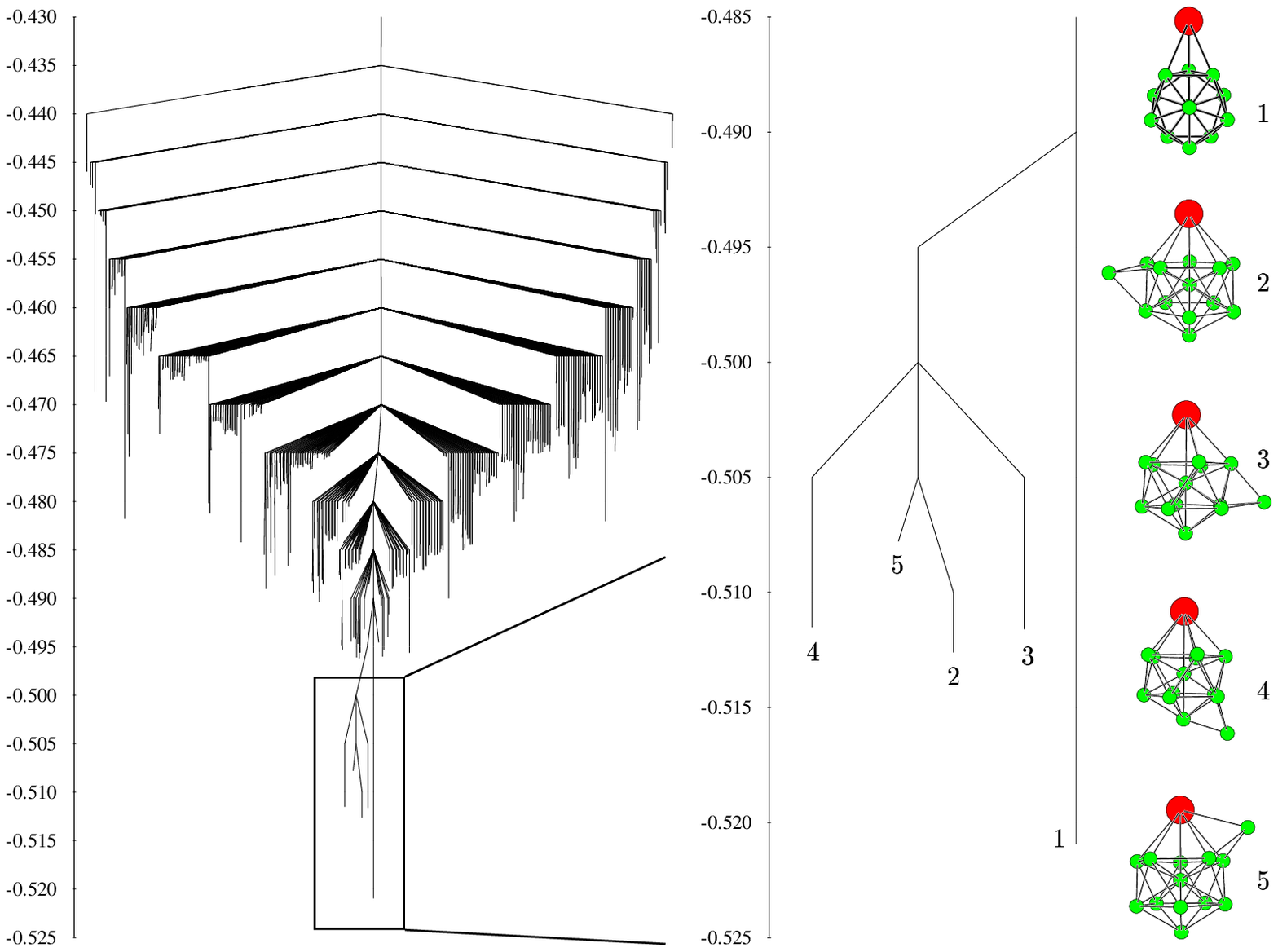}
\vfill}
\caption{Disconnectivity graph of the CaAr$_{13}$ cluster. The right panel
shows the lower part of the graph and the secondary funnel associated to
changing the substitutional site of calcium in the capped icosahedron
structure. All energies are in eV.}
\label{fig:graph13}
\end{figure*}
The    disconnectivity   graph    of   CaAr$_{13}$    is    shown   in
Fig.~\ref{fig:graph13}. The  `palm tree' character of  this graph also
suggests  that relatively fast  relaxation should  be possible  to the
global minimum.  However, the bottom  of the funnel is slightly rugged
due to the  presence of additional isomers with  respect to what would
be expected for Ar$_{14}$. The four extra structures found by changing
the substitutional  site of the  calcium atom create a  secondary side
funnel.  They are  close  in energy  to  the global  minimum, and  are
displayed in  Fig.~\ref{fig:graph13} next to  a more detailed  view of
the  corresponding  part of  the  graph.  The  ordering between  these
isomers  is  straightforwardly explained  by  the  weaker Ca--Ar  bond
compared  to  Ar$_2$. The  lowest-energy  barrier  between the  global
minimum  and the  next four  isomers lies  about  $2.19\times 10^{-3}$
eV/atom above isomer 1, which  is equivalent to a temperature of about
25\,K. Below this  temperature we expect the relaxation  to the global
minimum  to  become  significantly  slower  due  to  trapping  in  the
secondary funnel.

\begin{figure*}[htb]
\vbox to 12cm{
\includegraphics{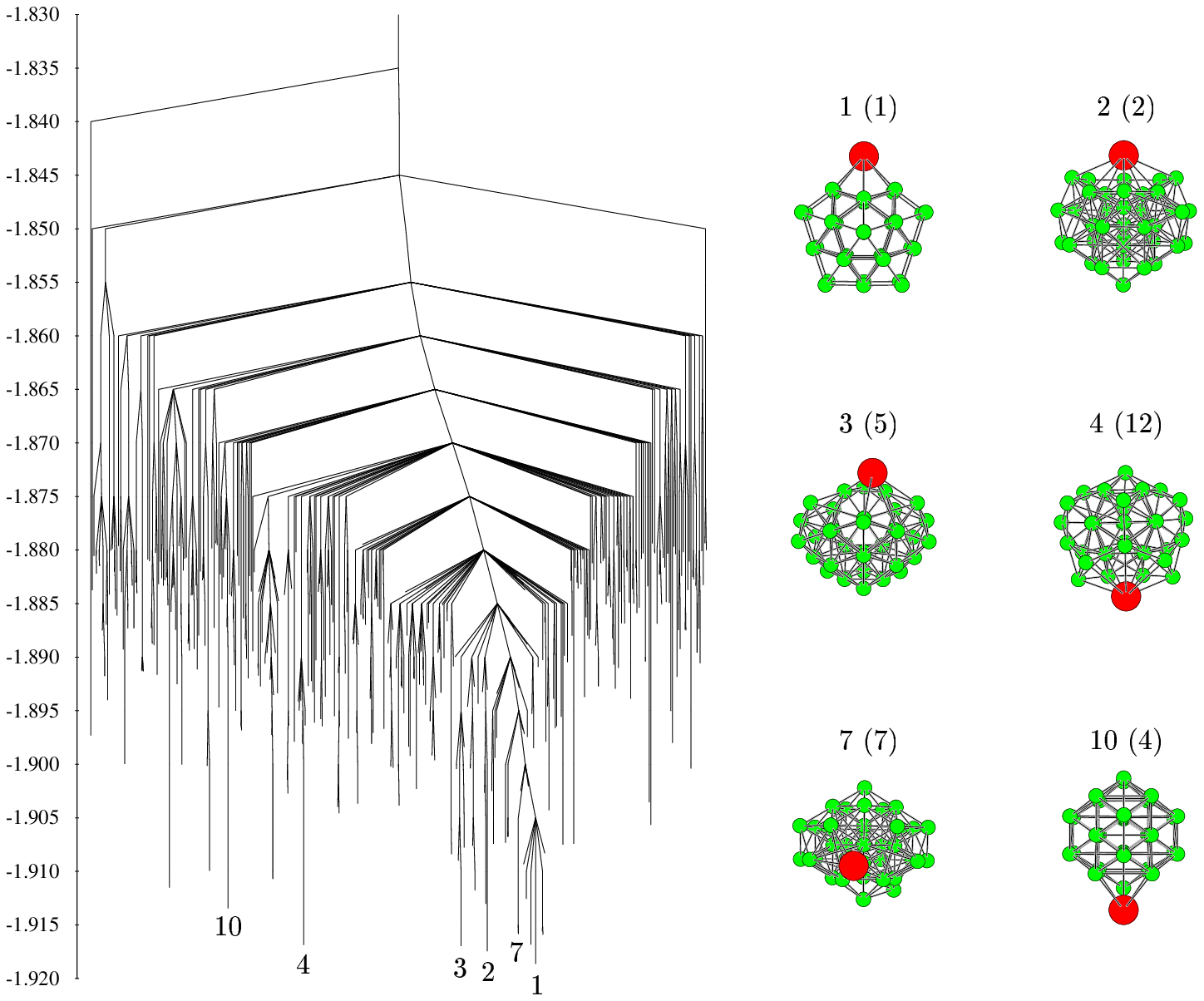}
\vfill}
\caption{Disconnectivity graph of the CaAr$_{37}$ cluster. Some isomers are
represented on the right side of the graph, with their rank in the quantum
(classical) regime. The energies are in eV.}
\label{fig:graph37}
\end{figure*}

Finally we consider the  larger CaAr$_{37}$ cluster. Starting from the
ten  lowest minima in  the database  obtained by  systematic quenching
along  a parallel  tempering Monte  Carlo  trajectory,\cite{paperI} we
were able  to establish twenty  connected discrete pathways  using the
method described in  reference \onlinecite{dps}. New low-energy minima
were discovered, two of them lower  than some in the initial set. Four
additional  pathways  were  then  constructed  in the  same  way.   As
mentioned above, these initial pathways  found may be far from optimal
in terms  of rates. The  typical initial path contained  30--40 minima
and  the  disconnectivity  graph  constructed just  from  the  initial
pathway  searches contained  731  distinct minima  and 820  transition
states.   It  is  shown  in Fig.~\ref{fig:graph37},  together  with  a
selection   of   low-energy  structures   chosen   from  the   initial
sample. Clearly this  graph does not show the  single funnel aspect of
the  previous   clusters,  and   resembles  more  the   `banyan  tree'
form.\cite{walesnat}

In fact, the energy landscape  of CaAr$_{37}$ is more complicated than
that of  Ar$_{38}$, which  only has two  main funnels  associated with
truncated  octahedral or  icosahedral  structures, respectively. It is
closer to the actual multiple funnel shape of the disconnectivity graph
in the ionic cluster (NaCl)$_{35}^-$.\cite{doyew99b} This
increased  complexity is  due  to  the larger  number  of isomers  and
possible  binding  sites  for  calcium,  and to  the  existence  of  a
particularly  stable  decahedral  isomer.  The  stabilization  of  the
anti-Mackay  icosahedral  structures  (isomer  4)  due  to  zero-point
effects is another complicating factor.\cite{jcpquantum}

The CaAr$_{37}$ cluster  exhibits several similarities with Ar$_{38}$,
such  as the  nonicosahedral  global minimum  and the  multiple-funnel
energy landscape. In thermal equilibrium, the low-temperature behavior
is essentially governed by  the decahedral and Mackay-type icosahedral
minima.\cite{paperI} The  truncated octahedral isomer  is not expected
to  play  a  major role  in  the  relaxation  dynamics, except  if  it
contributes significantly to the initial conditions.

From  Fig.~\ref{fig:graph37} we  see  that the  decahedral and  Mackay
icosahedra  isomers belong to  the same  superbasin at  relatively low
energies.   Hence we  expect to  find significantly  faster relaxation
than in  Ar$_{38}$, where the truncated octahedron  and the icosahedra
are  much farther  apart in  configuration space,  and separated  by a
relatively high barrier. However,  to obtain insight into the relative
stabilities  of the  isomers at  the temperature  where  relaxation is
simulated,  it may  be  helpful  to include  entropic  effects in  the
disconnectivity graphs.\cite{KrivovK02,EvansW03}  To achieve this goal
we  simply computed  the free  energies of  the minima  and transition
states   instead   of  potential   energy   minima,  replacing   $E_j$
($E_j^\alpha$)    by    $F_j=-k_BT\ln    Z_j$    ($F_j^\alpha=-k_BT\ln
Z_j^\alpha$),  where  the partition  functions  were  obtained in  the
harmonic         approximation,         and         taken         from
Eqn.~(\ref{eq:me5},\ref{eq:me6}).

The free-energy  disconnectivity graph of CaAr$_{37}$  at $T=20$\,K is
represented  in Fig.~\ref{fig:fegraph37}.  As can  be clearly  seen on
this  figure, the  most  stable  minimum at  this  temperature is  the
anti-Mackay icosahedron isomer (4), and the free energy barrier to the
decahedral  and  Mackay-icosahedral minima  seems  rather large.  This
change  in  the   most  stable  funnel  with  respect   to  the  $T=0$
disconnectivity  graph  should  hinder  relaxation, and  we  therefore
anticipate particularly slow dynamics.

\section{Relaxation to equilibrium: Time-dependent absorption spectra}
\label{sec:tdas}

We  now  turn  to  the  relaxation  properties  of  CaAr$_n$  clusters
resulting  from the  solution of  the  master equation.  Two types  of
relaxation have been considered,  either starting from $T=0$ (only the
quantum  global minimum  is occupied)  and  moving to  higher $T$,  or
starting  from  a  finite  temperature  equilibrium  distribution  and
changing to  lower $T$. The equilibrium  distributions were calculated
within the quantum harmonic superposition approach applied to the same
database  of minima  that was  used to  construct  the disconnectivity
graphs, without reweighting. For  CaAr$_{37}$, this procedure yields a
heat capacity  peak corresponding to  melting located at  about 25\,K,
slightly higher  than the one  obtained in the more  thorough previous
study.\cite{paperI} For the  other clusters the harmonic approximation
wthout reweighting performs rather well.
\begin{figure}[htb]
\vbox to 12cm{
\includegraphics{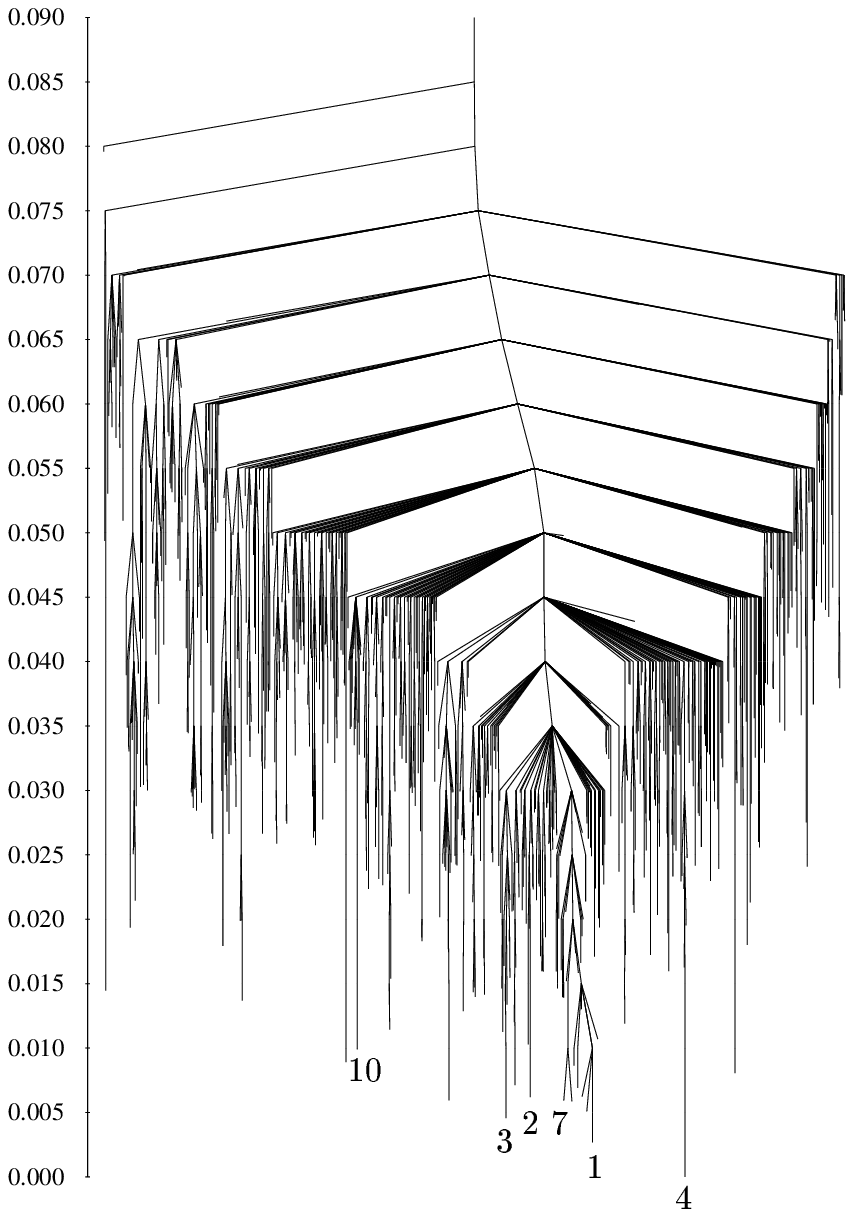}
\vfill}
\caption{Free-energy disconnectivity graph of the CaAr$_{37}$ cluster,
calculated at $T=20$\,K using harmonic quantum partition functions. The
isomers of Fig.~\protect\ref{fig:graph37} are represented. The
energies relative to the lowest free-energy minimum are in eV.}
\label{fig:fegraph37}
\end{figure}

Here we focus  on the photoabsorption spectrum as  a possible probe of
cluster structure in a dynamical context.  The absorption intensity of
the $4s^2\to 4s^14p^1$ excitation for calcium was calculated using the
Gaussian  theory  adapted and  extended  from  the  work of  Wadi  and
Pollak.\cite{paperI,wp}  The excited  state potential  energy surfaces
were modelled by a diatomics-in-molecules (DIM) Hamiltonian, which was
fully described in Ref.~\onlinecite{epjd}.

\begin{figure}[htb]
\vbox to 10.5cm{
\includegraphics{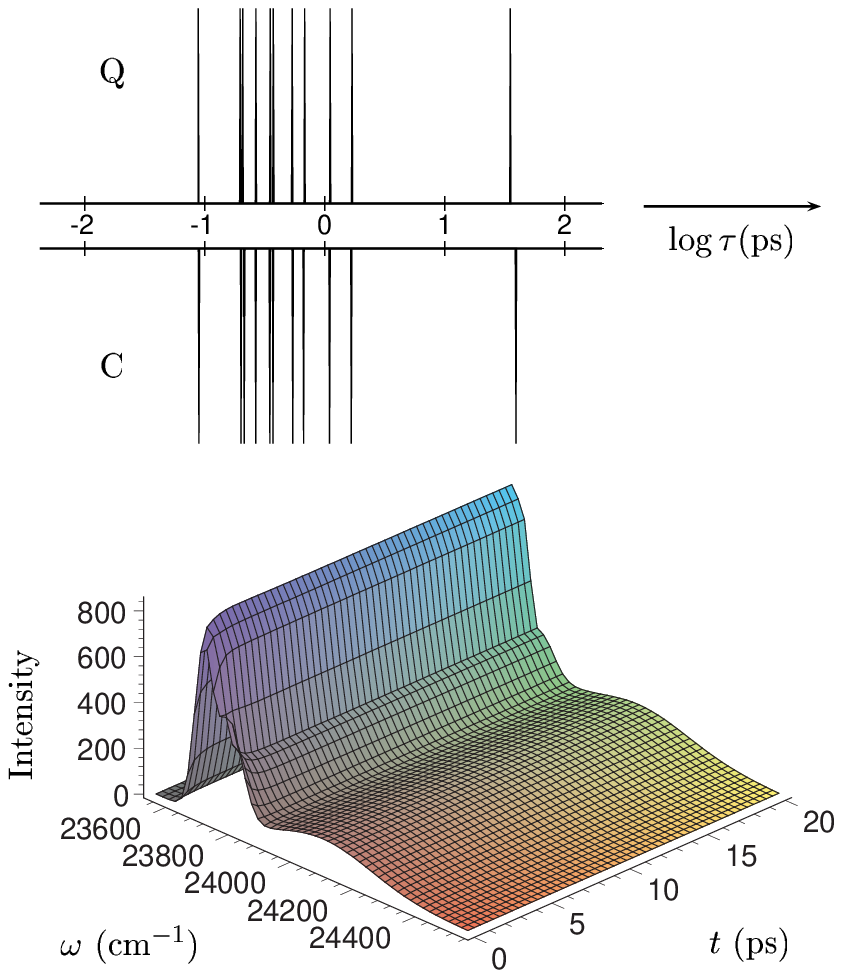}
\vfill}
\caption{Photoabsorption spectrum of CaAr$_6$ during the relaxation from the
$T=0$ distribution toward the $T=25$\,K equilibrium distribution. The discrete
spectra [in the quantum (Q) and classical (C) regimes] of the
characteristic times of the transition matrix ${\bf W}$ are represented on a
decimal
logarithmic scale, on the horizontal axis in the upper part of the figure.}
\label{fig:abs6}
\end{figure}
We  first  show  in  Fig.~\ref{fig:abs6}  the time  evolution  of  the
absorption  spectrum of  CaAr$_6$,  starting at  $t=0$  from the  zero
temperature  distribution, and relaxing  at $T=25$\,K,  slightly above
the  melting  point.\cite{paperI}  The  spectrum  remains  essentially
doubly-peaked, and  at very small times  $t<5$~ps the red  peak can be
separated into  two distinct lines characteristic of  the ground state
isomer. The  blue wing  of this peak  soon becomes a  shoulder.  These
results   correspond  to   fast   relaxation,  which   barely  has   a
spectroscopic signature. Such  simple dynamics is perfectly consistent
with  the  energy  landscape  of   this  cluster,  as  seen  from  its
disconnectivity graph. In the  upper panel of Fig.~\ref{fig:abs6}, the
characteristic time constants, $\tau_i$,  of the rate matrix ${\bf W}$
are  represented  on a  horizontal  axis,  for  both the  quantum  and
classical regimes.  They are calculated from  the nonzero eigenvalues,
$\lambda_i$, of  ${\bf W}$ as  $\tau_i = -1/\lambda_i$. At  25\,K, the
largest time constant  is a few tens of  picoseconds, which is roughly
in  agreement with  the observed  decay of  the  spectroscopic signal,
although the  experimental relaxationis faster, by about  one order of
magnitude,  than  the slowest  time  constant  $\tau_{\rm max}$.  Such
quantitative discrepancies are typical  of the errors that simple rate
theories  based upon  harmonic  densities of  states  are expected  to
introduce. Quantum effects are small in this system and the relaxation
kinetics for  CaAr$_{10}$ were found  to be very similar,  so detailed
discussion is omitted.

\begin{figure}[htb]
\vbox to 10.5cm{
\includegraphics{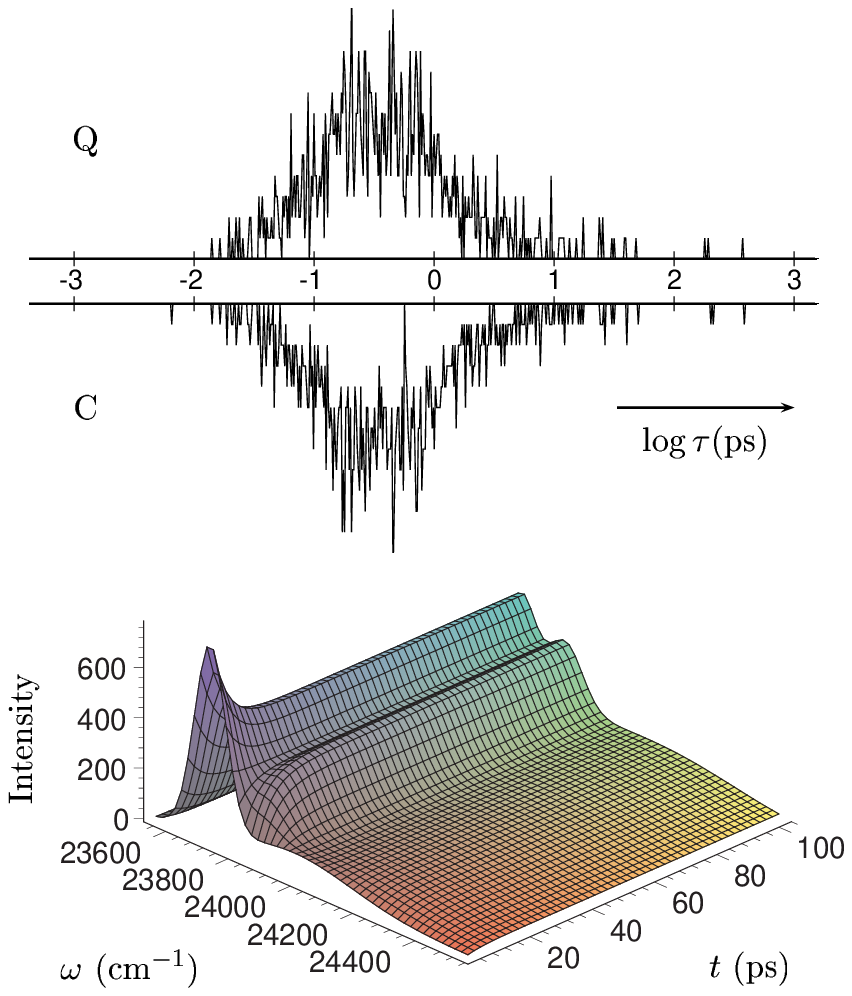}
\vfill}
\caption{Photoabsorption spectrum of CaAr$_{13}$ during the relaxation from the
$T=0$ distribution toward the $T=40$\,K equilibrium distribution. The continuous
distributions [in the quantum (Q) and classical (C) regimes] of the
characteristic times of the transition matrix ${\bf W}$ are represented on a
decimal
logarithmic scale, on the horizontal axis in the upper part of the figure.}
\label{fig:abs13a}
\end{figure}
Relaxation  from  a  zero   temperature  distribution  has  also  been
investigated in CaAr$_{13}$ for a temperature jump to $T=40$\,K, which
lies close to the melting  point of this cluster.\cite{paperI} At this
temperature,  the occupation  probability of  the secondary  funnel is
quite  large.  The  time-dependent spectrum  in  Fig.~\ref{fig:abs13a}
indeed exhibits a transition from  the ground state, identified by two
absorption peaks, to vibrationally  excited isomers and the three-peak
pattern   characteristic  of  calcium   located  in   the  icosahedral
shell.\cite{paperI}  Monitoring the  time-dependent signal  shows that
relaxation typically occurs on a  50~ps time scale, which is again one
order of  magnitude faster  than $\tau_{\rm max}$  in this  system, as
seen in the upper part of Fig.~\ref{fig:abs13a}.

\begin{figure*}[htb]
\vbox to 10.2cm{
\includegraphics{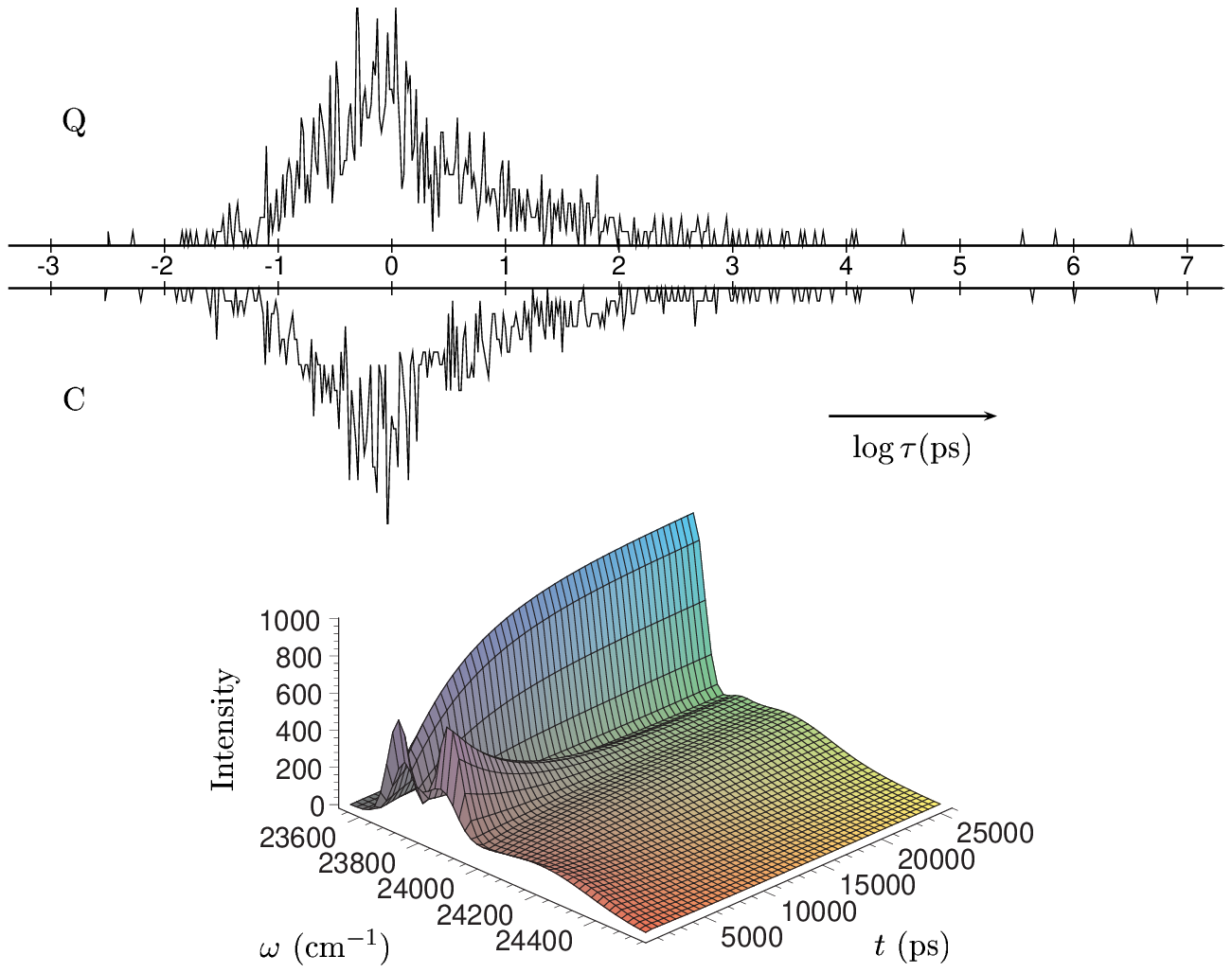}
\vfill}
\caption{Photoabsorption spectrum of CaAr$_{13}$ during the relaxation from the
$T=50$ distribution toward the $T=20$\,K equilibrium distribution. The
continuous distributions [in the quantum (Q) and classical (C) regimes] of the
characteristic times of the transition matrix ${\bf W}$ are represented on a
decimal
logarithmic scale, on the horizontal axis in the upper part of the figure.}
\label{fig:abs13b}
\end{figure*}
At $T=20$\,K, the  global minimum of CaAr$_{13}$ is  predicted to have
an equilibrium  occupation probability larger  than 90\%,\cite{paperI}
hence  the next  metastable  isomers play  only  a minor  role in  the
thermodynamic   properties.   Relaxation   from  a   high  temperature
distribution (50\,K) has also been simulated using the master equation
approach.   The  time-dependent  absorption  spectrum  is  plotted  in
Fig.~\ref{fig:abs13b}  along with  the distribution  of characteristic
decay  times from  the rate  matrix.  The variations  observed in  the
spectrum are in clear contrast  with the previous results.  During the
first  100\,ps,  the cluster  relaxes  from  a  broad distribution  of
isomers  towards  the  main  funnel  that corresponds  to  the  capped
icosahedron.  As indicated  by the  two-peak absorption  spectrum, the
secondary funnel has a larger occupation probability about 5\,ns after
the  dynamics has  started. The  cluster then  finally relaxes  to its
equilibrium   distribution  following  single   exponential  kinetics,
characterized by a rather large time constant of about $10^5$~ps. This
value  is still  one order  of  magnitude faster  than the  calculated
$\tau_{\rm  max}\approx 10^6$~ps,  represented  in the  upper part  of
Fig.~\ref{fig:abs13b}.  The two-step  relaxation process observed here
can  be easily  interpreted  thanks to  the  disconnectivity graph  in
Fig.~\ref{fig:graph13}.  The secondary  funnel  constitutes a  kinetic
trap, resulting in  a significant slowing down of  the dynamics. While
the cluster easily finds its way into this secondary funnel, escape to
the  global minimum  is much  slower due  to the  rather  large energy
barrier.   In  agreement  with  previous  work,  this  multiple-funnel
landscape    exhibits   a    separation   of    timescales    in   its
dynamics.\cite{walesnat,walesacp}

\begin{figure*}[htb]
\vbox to 10cm{
\includegraphics{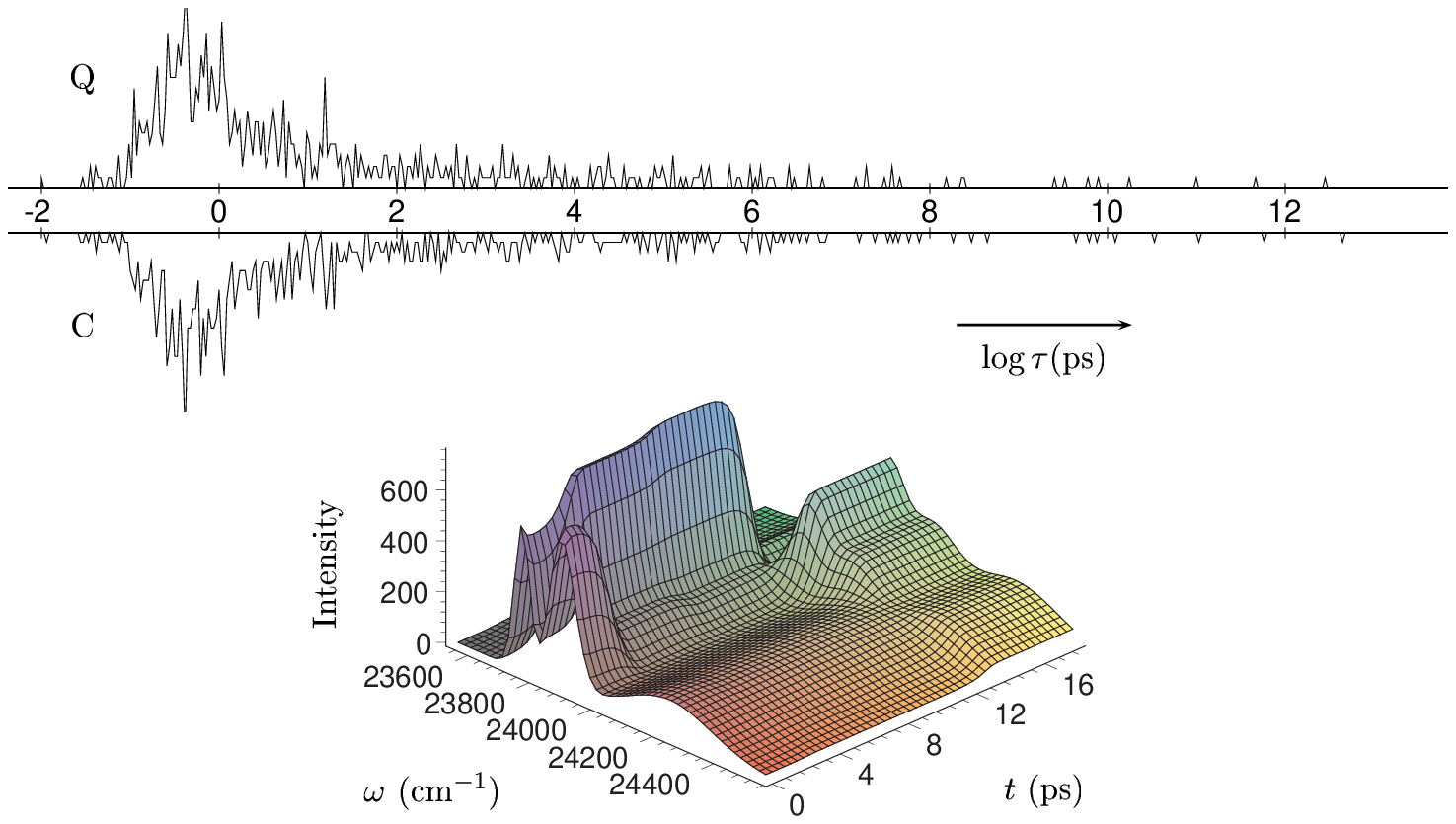}
\vfill}
\caption{Photoabsorption spectrum of CaAr$_{37}$ during the relaxation from the
$T=0$ distribution toward the $T=20$\,K equilibrium distribution. The continuous
distributions [in the quantum (Q) and classical (C) regimes] of the
characteristic times of the transition matrix ${\bf W}$ are represented on a
decimal
logarithmic scale, on the horizontal axis in the upper part of the figure.}
\label{fig:abs37}
\end{figure*}
We finally  discuss the case of  CaAr$_{37}$, for which  we solved the
master  equation  at  $T=20$\,K,   starting  from  the  $T=0$  initial
distribution.  The magnitude  of  the characteristic  decay times,  as
computed  from the  eigenvalues  of the  rate  matrix, are  remarkably
large, reaching values close  to one second. The calculated absorption
spectrum is shown  in Fig.~\ref{fig:abs37} as a function  of time. The
choice  of a logarithmic  scale emphasizes  the different  time scales
present in this  system. The multiple funnels of  the energy landscape
are  in competition below  20\,K, and  the cluster  undergoes multiple
transitions before  reaching equilibrium. As we deduced  from the free
energy  disconnectivity graph in  Fig.~\ref{fig:fegraph37}, relaxation
from 0 to 20\,K in this system requires the cluster to escape from its
initial  decahedral/icosahedral  funnel.  At  least  four  significant
changes  in behaviour  are detected  through  absorption spectroscopy,
after $10^2$,  $10^6$, $10^{10}$, and  $10^{12}$~ps, respectively. The
transient regime thus  appears particularly slow, in a  similar way to
the  dynamics observed in  Ar$_{38}$.\cite{ar38miller} The  spectra in
Fig.~\ref{fig:abs37} show that some intermediate minima, which are too
high in energy to be  significantly occupied at equilibrium, may still
contribute to the spectrum as transient species with a long lifetime.

For CaAr$_{37}$ the set  of connected minima was initially constructed
in a rather artificial manner  by means of direct pathway searches. As
mentioned  in the previous  section, these  initial paths  are usually
rather   long  and   correspond  to   high  barriers.   Therefore  the
interconversion  rates  predicted  using  this database  are  probably
unrealistic   and   much  too   low.   The   discrete  path   sampling
technique\cite{dps} was therefore used to optimize the pathway between
the  two lowest  decahedral and  Mackay icosahedral  minima.  We found
pathways  containing fewer than  seven minima  (including the  two end
points) with  a rate constant  about eight orders of  magnitude larger
than that  of the  initial 45-step pathway.  This improvement  is even
more impressive  than the one found  for Ar$_{38}$.\cite{dps} However,
in  CaAr$_{37}$ the  distance between  the two  minima in  question is
rather shorter, because they belong to the same funnel.

We did not attempt to optimize the 23 other initial pathways using the
discrete path sampling method,
since  the effect  of  refining the  interconversion  pathways is  not
expected to affect  the complex variation with time  of the absorption
intensity.

\section{Conclusions}
\label{sec:ccl}

The  complexity  of the  energy  landscapes  of  CaAr$_n$ clusters  is
reflected not only on their  equilibrium properties, but also in their
relaxation  kinetics.   By  employing  the master  equation  approach,
quantum corrected harmonic partition functions and the Gaussian theory
of  absorption,\cite{wp} we have  investigated the  time-dependence of
the photoabsorption  intensity. The clusters were chosen  to provide a
selection  of various  finite-size effects.  Our main  result  is that
spectroscopy may be able to probe the isomerization of a given cluster
in  real time,  and  provide estimates  of  the interconversion  rates
themselves. We have found  evidence that CaAr$_n$ clusters can display
simple kinetics  (as in  CaAr$_6$ or CaAr$_{10}$),  two-state kinetics
and  trapping in  auxiliary funnels  (aAr$_{13}$), or  more intricate
kinetics  and multiple  transient  regimes (CaAr$_{37}$),  all with  a
distinct spectroscopic signature. 
experiments?

A number  of approximations have  been used to achieve  these results,
including  the harmonic  superposition approximation  using incomplete
samples  of  minima  for  the  larger clusters.  The  master  equation
approach  also contains  some assumptions,  such as  local equilibrium
inside  each  basin and  Markovian  dynamics.   Finally, the  Gaussian
theory of  absorption also  contains some approximations.   Because of
these  restrictions, our investigation  may be  semi-quantitative with
respect  to actual  CaAr$_n$ clusters,  even before  we allow  for the
approximate nature of  the Hamiltonians used to model  the ground- and
excited-state potential  energy surfaces.\cite{epjd} Nevertheless, the
present  study   provides  the  next  step  beyond   the  analysis  of
equilibrium properties  discussed in our  previous paper.\cite{paperI}
We believe that it further supports the need for future experiments on
size-selected, trapped clusters studied via spectroscopic techniques.

\end{document}